# Translation-Invariant Renormalizable Noncommutative Chern-Simons Theory

Kaddour Chelabi[a], Manfred Schweda[b] and Smain Kouadik[c]

a- Phsyics Department, University of New Brunswick
Fredericton, New Brunswick (Canada).
b-Insitute for Theoretical Physics, Vienna University of Technology
Wiedner Hauptstrasse 8-10, A-1040 Vienna (Austria).
c-Laboratory of mechanics, physics and mathematical modeling
Faculty of Science and Technology Medea University 26000. Medea (Algeria).
E-mail: kaddourchelabi@unb.ca , mschweda@tph.tuwien.ac.at , skouadik@hotmail.com

## Abstract

In this paper we show the renormalizability of the translation invariant noncommutative Chern-Simons theory, motivated by the work done on noncommutative scalar field theory [06]. We add a new term to the bilinear part of the action. In addition, we prove, the finiteness of the theory at one- and two-loop level despite this modification. Finally we perform the one-loop two point functions of the gluon contribution.

# Introduction:

In the last fifteen years, a lot of work has been devoted to the study of field theories on noncommutative Moyal space, that is space-time coordinates should no longer commute at the Planck scale where gravity should be quantized. The main motivation for these theories arises from string theory[01], after the general mathematical framework developed by A.Connes and others[02]. Renormalization is the center interest of field theory, however, it was found that noncommutative theories are not renormalizable because of a surprising phenomenon called UV/IR[03], which means that the theory still has infinities. Fortunately, this is new kind of divergence was solved by adding a properly chosen term to the theory, which has proved to provide a cure to these divergences, and make them renormalizable [04,05,06].

Noncommutative Chern-Simons as a noncommutative gauge theory belongs to the UV/IR-mixing theories [07], in addition, it was shown as well the finiteness and the independence of the deformation parameter $\theta^{\mu\nu}$ at least at one loop level. Our aim in this work is to get rid of this divergence by modifying the kinetic term of the action. Our paper is organized as follows: After introducing the Chern-Simons action with respect to a non-abelian $U(N)$-gauge group on noncommutative $R^3$, adding a non-local counter-term to the bilinear part. We perform the perturbative analysis of this modified model, going through the Faddeev-Popov quantization, the Feynman rules and the finiteness of the theory at one- and two-loop level in section four. At the end, section three is devoted to explicit calculation of one-loop two-point function of the gluon contribution.

## 1-The New Version of Noncommutative Chern-Simons Theory:

The modified noncommutative Chern-Simons action is given as follows:

$$S = \frac{\kappa}{4\pi} \int d^3x \varepsilon^{\mu\nu\rho} Tr\left( A_\mu \tilde{\nabla}_\nu A_\rho + \frac{2i}{3} A_\mu * A_\nu * A_\rho \right), \qquad (1)$$

where:

$$\tilde{\nabla}_\mu = \left(1 + \frac{1}{\theta^2 \Box^2}\right)\partial_\mu \qquad (2)$$

and the star-product is simply the Moyal-product

$$(f * g)(x) = f(x) e^{i \overleftarrow{\partial}_\mu \theta^{\mu\nu} \overrightarrow{\partial}_\nu} g(x) = \int \frac{d^3k}{(2\pi)^{3/2}} \frac{d^3p}{(2\pi)^{3/2}} e^{i(k_\mu + p_\mu)x^\mu} e^{-\frac{i}{2}\theta^{\mu\nu} k_\mu p_\nu} \tilde{f}(k)\tilde{g}(p), \qquad (2)$$

which gives the three-dimensional noncommutative coordinates commutation relations

$$\left[x^\mu, x^\nu\right]_* = i\theta \theta^{\mu\nu}. \qquad (4)$$

where

$$\theta^{\mu\nu} = \begin{pmatrix} 0 & 1 & 1 \\ -1 & 0 & 1 \\ -1 & -1 & 0 \end{pmatrix} \quad , \quad \theta \in R, \quad (5)$$

are the antisymmetric and real parameters of dimension length squared with full rank.

we notice here that the dynamical field is the gauge potential $A_\mu = A_\mu^a T^a$ the generators of the gauge group $G = U(N)$, normalized to $Tr(T^a T^b) = 2\delta^{ab}$. The used group algebra tools in the adjoint representation are:

$$\begin{cases} Tr(T^a T^b T^c) = 2(if^{abc} + d^{abc}) \quad , \quad \{T^a, T^b\} = 2d^{abc} T^c \quad , \quad [T^a, T^b] = 2if^{abc} T^c \\ d^{abc} d^{dbc} = N(\delta^{ad} + \delta^{a0}\delta^{d0}) \quad , \quad f^{abc} f^{dbc} = N(\delta^{ad} - \delta^{a0}\delta^{d0}) \quad , \end{cases} \quad (6)$$

where $f^{abc}, d^{abc}$ are the antisymmetric and the symmetric structure constants of the group respectively. The action is invariant under the modified-gauge transformations

$$\tilde{\delta} A_\mu \equiv \tilde{D}_\mu \lambda = \tilde{\nabla}_\mu \lambda + i[\lambda, A_\mu]_*, \quad (7)$$

with $\lambda$ the Lie-algebra valued gauge parameter. The filed strength becomes

$$\tilde{F}_{\mu\nu} = \tilde{\nabla}_\mu A_\nu - \tilde{\nabla}_\nu A_\mu - i[A_\mu, A_\nu]_*. \quad (8)$$

The model (1), inspired by the work done in [06], consists of adding a non-local counterterm for the quadratic IR divergence of the original model. This procedure provides a solution for the UV/IR-mixing problem while maintaining translation invariance. In the coming sections, we perform the perturbative analysis by following the standard scheme of path integral quantization to establish the Feynman rules.

## 2-Perturbative NC C-S Theory:

## 2-1- The Total Action:

We use the standard Faddeev-Popov procedure to perform the quantization, in Euclidean space the total action becomes

$$S_{NCCS} = S_{cl} + S_{gf} + S_{ext}$$

$$= -\int d^3x \frac{1}{2}\varepsilon^{\mu\nu\rho}Tr\left(A_\mu \tilde{\nabla}_\nu A_\rho + \frac{2i}{3}A_\mu * A_\nu * A_\rho\right)$$

$$+ \int d^3x Tr\left(B * \tilde{\nabla}_\mu A^\mu - \bar{c}\tilde{\nabla}_\mu \tilde{D}^\mu c\right)$$

$$+ \int d^3x Tr\left(\rho_\mu * \tilde{D}^\mu c + i\sigma * c * c\right), \tag{9}$$

where $S_{cl}$ represents the modified classical action(1), and $S_{gf}$ is the gauge fixing action, and $S_{ext}$ the external source action. B is the Lagrange multiplier, $c$ and $\bar{c}$ are the ghost and anti-ghost fields respectively. The action (9) is invariant under the BRST-transformations

$$\begin{cases} sA_\mu = \tilde{D}_\mu c, \\ sc = ic * c, \\ s\bar{c} = B, \\ sB = 0. \end{cases} \tag{10}$$

The BRS transformations are nilpotent nonlinear and supersymmetric. Moreover, they are rather useful to derive Ward-identities between the Green functions of the theory. Actually, they represent the role of gauge invariance for the gauge fixed theory. In other words, the gauge symmetry of the theory implies certain relations among the Green functions of theory.

## 2-2-Feynman Rules:

### 2-2-1-The Propagators:

From the quadratic term of relation (9) one gets:

1- the $AA$-propagator:

$$\Delta^{AA,ab}_{\mu\nu} = \frac{i\delta^{ab}}{(2\pi)^{3/2}} \frac{\varepsilon_{\mu\lambda\nu}\partial^\lambda}{\left(1+\frac{1}{\theta^2\Box^2}\right)\Box}. \tag{11}$$

2- the $c\bar{c}$- propagator:

$$\Xi^{c\bar{c},ab} = \frac{\delta^{ab}}{(2\pi)^{3/2}} \frac{1}{\left(1+\frac{1}{\theta^2\Box^2}\right)\Box}. \tag{12}$$

3-the *AB*-propagator:

$$\Xi_\mu^{AB,ab} = \frac{-i\delta^{ab}}{(2\pi)^{3/2}} \frac{\partial_\mu}{\left(1+\frac{1}{\theta^2\Box^2}\right)\Box}.$$  (13)

## 2-2-2-The Vertices :

From the interaction part of the action (9) we get in momentum space :

1- The *AAA*-vertex:

$$V_{\mu\nu\rho}^{AAA,abc}(p_1,p_2,p_3) = \frac{1}{(2\pi)^{3/3}} \varepsilon_{\mu\nu\rho} \Omega_\theta^{abc}(p_2,p_3),$$  (14)

2- the $cA\bar{c}$-vertex:

$$V_\mu^{cA\bar{c},abc}(p_1,p_2,p_3) = \frac{-i}{(2\pi)^{3/3}} \left(1+\frac{1}{\theta^2 p^4}\right) p_{3\mu} \Omega_\theta^{abc}(p_2,p_3),$$  (15)

where

$$\Omega_\theta^{abc}(p_2,p_3) = 4\left[f^{abc}\cos\left(\frac{1}{2}p_2\times p_3\right) - d^{abc}\sin\left(\frac{1}{2}p_2\times p_3\right)\right]$$

and $p_2 \times p_3 = p_{2\mu}\theta^{\mu\nu}p_{3\nu} = p_2\tilde{p}_3 = -\tilde{p}_2 p_3$.

We notice here that the Green functions(the propagators and the vertices) are related by the Ward-identities as follows:

$$i\varepsilon_{\mu\nu\rho}\left(1+\frac{1}{\theta^2 p^4}\right) p^\rho \Xi^{c\bar{c}}(p) = \Delta_{\mu\nu}^{AA,ab}(p),$$  (16)

$$i\left(1+\frac{1}{\theta^2 p^4}\right) p_3^\nu \varepsilon_{\mu\nu\tau} V_{\lambda\sigma\tau}^{AAA,abc}(p_1,p_2,p_3) = \delta_{\mu\sigma} V_\lambda^{cA\bar{c}}(p_1,p_2,p_3) - \delta_{\mu\lambda} V_\sigma^{cA\bar{c},abc}(p_1,p_2,p_3).$$  (17)

Knowing that are always an output of a symmetry. The last equations are obtained from the linear VSUSY symmetry:

$$\begin{cases} \delta_\nu A_\tau = \varepsilon_{\mu\nu\tau}\tilde{\nabla}_\mu\bar{c}, \\ \delta_\nu c = -A_\nu, \\ \delta_\nu B = -\tilde{\nabla}_\nu\bar{c}, \\ \delta_\nu\bar{c} = 0. \end{cases}$$  (18)

for more details see [07] and references within.

## 2-3-The finiteness at one loop:

### 2-3-1- The two-point functions:

The gluon self-energy has three contributions, one from the gluon loop and the other two from the ghost loop (two directions) Figure-1- (the dashed line represent the gluons, and the continuous line represents the ghost). Using the Feynman rules relations (11), (12), (14) and (15), we may write the two point function of the gluon contributions

$$\Pi^{ab,AA}_{\mu\nu,gluon}(p,-p) = \frac{16N}{(2\pi)^6}\int \frac{d^3k}{(2\pi)^{3/2}} \frac{k_\mu(k+p)_\nu + k_\mu(k+p)_\nu}{\left[(k+p)^2 + \frac{1}{\theta^2(k+p)^2}\right]\left[k^2 + \frac{1}{\theta^2 k^2}\right]} [\delta^{ab} - \delta^{a0}\delta^{b0}\cos(p\times k)]. \quad (19)$$

Besides, the one-loop tow-point functions of the two ghost contributions may be written as follows

$$\Pi^{ab,AA}_{\mu\nu,ghost}(p,-p) = -\frac{16N}{(2\pi)^6}\int \frac{d^3k}{(2\pi)^{3/2}} \frac{k_\mu(k+p)_\nu + k_\nu(k+p)_\mu}{\left[(k+p)^2 + \frac{1}{\theta^2(k+p)^2}\right]\left[k^2 + \frac{1}{\theta^2 k^2}\right]} [\delta^{ab} - \delta^{a0}\delta^{b0}\cos(p\times k)]. \quad (20)$$

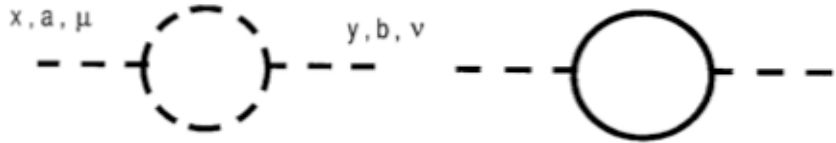

Figure-1-: gluon self-energy.

we can see that the sum of the whole contributions is zero; including one gluon and two ghost contributions(two senses).

### 2-3-2-The three-point functions:

For the three-gluon-vertex contributions we have the gluon one and the ghost one Figure-2-,

$$\Gamma^{abc}_{\mu\nu\rho}(p_1,p_2,p_3) = -i\int \frac{d^3k}{(2\pi)^{3/2}} \frac{k_\mu(k-p_2)_\nu(k+p_1)_\rho \Omega^{adl}_\theta(-p_1,k)\Omega^{blj}_\theta(k,p_2)\Omega^{cjd}_\theta(-p_3,k+p_1)}{\left[(k+p_1)^2 + \frac{1}{\theta^2(k+p_1)^2}\right]\left[(k-p_2)^2 + \frac{1}{\theta^2(k-p_2)^2}\right]\left[k^2 + \frac{1}{\theta^2 k^2}\right]}. \quad (21.\ a)$$

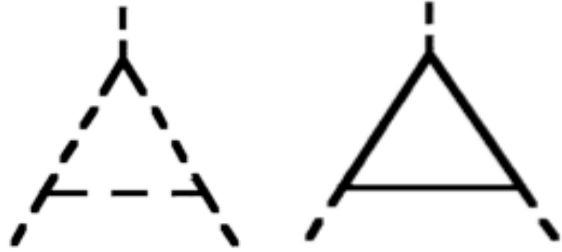

Figure -2-

The relation is proportional to completely antisymmetric product, where we used Schwinger parameterization to show that it vanishes. In addition, we have also the ghost-gluon-antighost contributions (Figure-3-), which can be written in function of relation (21),

$$\Gamma_\mu^{abc}(p_1,p_2,p_3)_{1+2} = \frac{-i}{3!}\left(1+\frac{1}{\theta^2 p^4}\right) p_{3\mu} \varepsilon_{\nu\rho\sigma} \Gamma_{\nu\rho\sigma}^{abc}(p_1,p_2,p_3) \ . \tag{21. b}$$

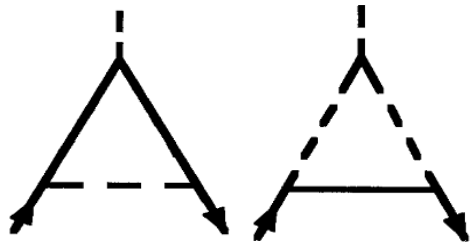

Figure-3-

The above results show that the theory is one-loop finite.

## 2-4-Finiteness at two-loop level:

Figure-4-a, represents the whole possible diagrams of the two-point function contributions at two-loop level, knowing that each diagram contains a combination of whether the gluon or the ghost contributions relations (19,20). Figure-4-b, show the possible contributions of the three-point functions at two-loop level, which vanish since it is a combination of the vertex functions relations(21.a.b). We see that the theory is finite at two-loop level. For more details one can see[09].

Finally, we say that the noncommutative Chern-Simons with the correction term –model (9)- belongs to the finite theories, and the whole radiative corrections vanish as well as the commutative case[10].

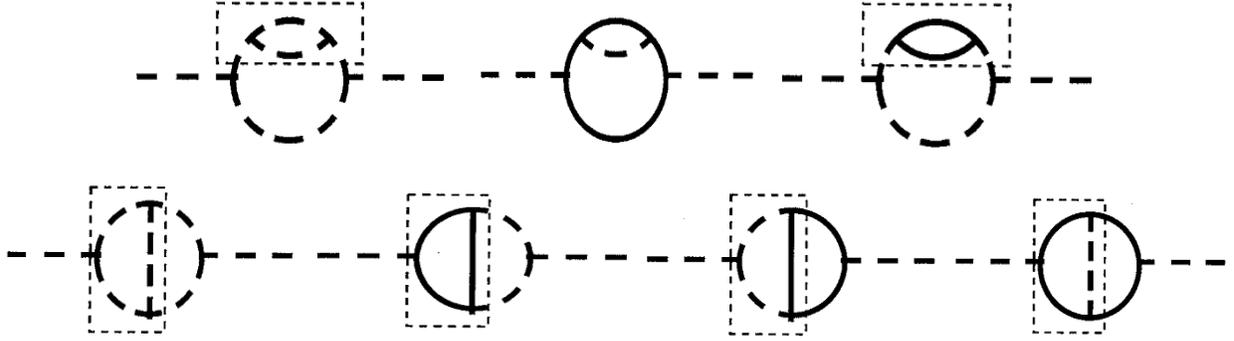

Figure-4-a-

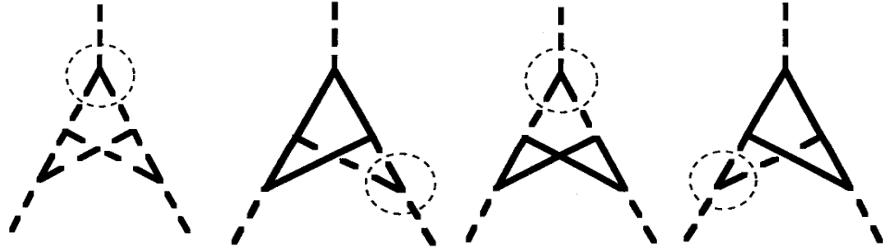

figure-4-b-

## 3- The One-Loop Two-Point Function:

In the following we are going to carry out explicitly the gluon contribution (Figure-1-), of the gluon self energy relation (19). The technique we will use is already used in [08]. Using the Schwinger parameterization. Simplifying the denominator as follows

$$\begin{cases} \dfrac{1}{\left[k^2 + \dfrac{1}{\theta^2 k^2}\right]} = \dfrac{1}{2}\sum_{\xi=\pm i}\dfrac{1}{k^2 + \xi\theta}, \\ \dfrac{1}{\left[(k+p)^2 + \dfrac{1}{\theta^2(k+p)^2}\right]} = \dfrac{1}{2}\sum_{\xi=\pm i}\dfrac{1}{(k+p)^2 + \xi\theta}. \end{cases} \quad (22.\,a)$$

As usually in using Schwinger parameter one gets:

$$\begin{cases} \dfrac{1}{(k+p)^2+\xi\theta} = \displaystyle\int_0^\infty d\alpha\ e^{-\alpha\left[(k+p)^2+\xi\theta\right]}, \\ \dfrac{1}{k^2+\xi\theta} = \displaystyle\int_0^\infty d\beta\ e^{-\beta\left[k^2+\xi\theta\right]}. \end{cases} \quad (22.\ b)$$

Substituting in (19), we get

$$I_{\mu\nu} = \frac{16N}{(2\pi)^6}\frac{1}{4}\sum_{\xi=\pm i\eta\pm i}\int\frac{d^3k\,d\alpha\,d\beta}{(2\pi)^{3/2}}\left[k_\mu(k+p)_\nu + k_\nu(k+p)_\mu\right]e^{-\alpha\left[(k+p)^2+\xi\theta\right]-\beta\left[k^2+\xi\theta\right]+\eta k\times p}, \quad (23)$$

where $\eta=\pm i$ represent the $\delta^{a0}\delta^{b0}$-term and $\eta=0$ is the $\delta^{ab}$-term. After simplifying the integral and carrying out the Gaussian integral, we end up with

$$I_{\mu\nu}(\eta=0) = \frac{1}{2^{3/2}}\int_0^\infty d\alpha\,d\beta\,\frac{1}{(\alpha+\beta)^{5/2}}\left\{\delta_{\mu\nu} - \frac{2\alpha\beta}{(\alpha+\beta)}p_\mu p_\nu\right\}e^{-\frac{\alpha\beta}{(\alpha+\beta)}p^2-(\alpha+\beta)\xi\theta}, \quad (24)$$

$$I_{\mu\nu}(\eta=\pm i) = \frac{1}{2^{1/2}}\int_0^\infty d\alpha\,d\beta\,\frac{1}{(\alpha+\beta)^{5/2}}\left\{\delta_{\mu\nu} - \frac{2\alpha\beta}{(\alpha+\beta)}p_\mu p_\nu - \frac{1}{2(\alpha+\beta)}\tilde{p}_\mu\tilde{p}_\nu\right\}e^{-\frac{\alpha\beta}{(\alpha+\beta)}p^2 - \frac{1}{4(\alpha+\beta)}\tilde{p}^2-(\alpha+\beta)\xi\theta}, \quad (25)$$

where there is a sum over $\xi=\pm i$. Now, in order to perform these integrals we use a change of variable from $(\alpha,\beta)$ to $(\rho,t)$,

$$\begin{cases} \alpha = t\lambda, \\ \beta = (1-t)\lambda, \\ \rho = t(1-t)p^2\lambda. \end{cases} \quad (26)$$

Substituting these new variables in (24) and (25), with some simplifications, we get for (24)

$$I_{\mu\nu}(\eta=0) = -\sqrt{\frac{\pi}{2}}\left\{\delta_{\mu\nu}\int_0^\infty dt\left[t(1-t)p^2+\xi\theta\right]^{1/2} + p_\mu p_\nu\int_0^\infty dt\,\frac{t(1-t)}{\left[t(1-t)p^2+\xi\theta\right]^{1/2}}\right\}. \quad (27)$$

The integral (25) which contains three terms become

$$\begin{cases} I^1_{\mu\nu}(\pm i) = -\dfrac{2\delta_{\mu\nu}}{(\tilde{p}^2)^{1/4}} \int_0^1 dt \left[t(1-t)p^2 + \xi\theta\right]^{1/4} K_{-1/2}(z) & (28) \\[2mm] I^2_{\mu\nu}(\pm i) = -2p_\mu p_\nu (\tilde{p}^2)^{1/4} \int_0^1 dt \dfrac{t(1-t)}{\left[t(1-t)p^2 + \xi\theta\right]^{1/4}} K_{1/2}(z) & (29) \\[2mm] I^3_{\mu\nu}(\pm i) = -2\dfrac{\tilde{p}_\mu \tilde{p}_\nu}{(\tilde{p}^2)^{3/2}} \int_0^1 dt \left[t(1-t)p^2 + \xi\theta\right]^{3/4} K_{-3/2}(z) & (30) \end{cases}$$

where $z = \sqrt{\tilde{p}^2\left[t(1-t)p^2 + \xi\theta\right]}$ and $K_n(z)$ are the modified Bessel functions, which are given as follows

$$K_{1/2}(z) = K_{-1/2}(z) = \left(\dfrac{\pi}{2z}\right)^{1/2} e^{-z} \qquad (31)$$

$$K_{j-1}(z) - K_{j+1}(z) = \dfrac{-2j}{z} K_j(z) \qquad (32)$$

We notice here that, if we carry out the integral (27), with taking the case $\xi = 0$ or $\theta = 0$, we get

$$I^{com}_{\mu\nu}(0) = -\dfrac{\sqrt{2\pi}}{3}\left\{\sqrt{p^2}\delta_{\mu\nu} + \dfrac{p_\mu p_\nu}{\sqrt{p^2}}\right\}, \qquad (33)$$

which represents the commutative case, regularized by the Pauli-Villars and Higher derivative techniques [10].

Finally, taking the limit $p = 0$, of the integrals (27),(28),(29) and (30), we have

$$I_{\mu\nu}(\eta = 0) = -\sqrt{\dfrac{\pi}{2}}\left\{\delta_{\mu\nu}\left[c_1 + c_2 p^2 + O(p^4)\right] + p_\mu p_\nu\left[c_3 + c_4 p^2 + O(p^4)\right]\right\}, \qquad (34)$$

$$\begin{cases} I^1_{\mu\nu}(\pm i) = -\sqrt{2\pi}\,\dfrac{\delta_{\mu\nu}}{(\tilde{p}^2)^{1/2}}\left[c_7 + c_8 p^2 + O(p^4)\right], \\[2mm] I^2_{\mu\nu}(\pm i) = -\sqrt{\dfrac{\pi}{2}} p_\mu p_\nu\left[c_7 + c_8 p^2 + O(p^4)\right], & (35) \\[2mm] I^3_{\mu\nu}(\pm i) = -\sqrt{2\pi}\,\dfrac{\tilde{p}_\mu \tilde{p}_\nu}{\tilde{p}^{3/2}}\left[c_9 + c_{10} p^2 + O(p^4)\right], \end{cases}$$

The first and the third term of equ.(35) show IR-divergences but these singularities are canceled by the two ghost one-loop contributions as one can see from the statement of Sect.2-3-1.

*- and for the limit $p \to \infty$:

$$I_{\mu\nu}(\eta=0) = -\sqrt{\frac{\pi}{2}}\left\{\delta_{\mu\nu}\left[c'_1 p + \frac{c'_2}{p} + \frac{c'_3}{p^3} + O\left(\frac{1}{p^5}\right)\right] + p_\mu p_\nu\left[\frac{c'_3}{p} + \frac{c'_4}{p^3} + O\left(\frac{1}{p^5}\right)\right]\right\}, \quad (36)$$

$$\begin{cases} I^1_{\mu\nu}(\pm i) = -\sqrt{2\pi}\,\dfrac{\delta_{\mu\nu}}{(\tilde{p}^2)^{1/2}}\left[\dfrac{c'_6}{p} + \dfrac{c'_7}{p^2} + O\left(\dfrac{1}{p^3}\right)\right], \\[2ex] I^2_{\mu\nu}(\pm i) = -\sqrt{\dfrac{\pi}{2}}\,p_\mu p_\nu\left[\dfrac{c'_8}{p} + \dfrac{c'_9}{p^2} + O\left(\dfrac{1}{p^3}\right)\right], \\[2ex] I^3_{\mu\nu}(\pm i) = -\sqrt{2\pi}\,\dfrac{\tilde{p}_\mu \tilde{p}_\nu}{(\tilde{p}^2)^{9/4}}\left[c_{10} + c_{11}p^2 + \dfrac{c'_{12}}{p} + \dfrac{c'_{13}}{p^2} + O\left(\dfrac{1}{p^3}\right)\right], \end{cases} \quad (37)$$

where c's are constants. We notice that at large momentum scale, the results show no UV-divergences.

**4-Conclusion and Outlook:**

In this work, the noncommutative Chern-Simons theory with the extra term, model (9), keeps having the property of the finiteness at one- and two-loop level, which is the same property as the commutative case. Moreover, since any noncommutative theory contains the new type of divergence UV/IR-mixing, the regularization of this issue is still not achieved as shown in equation (34), which contains IR-singularity. However, as one can see from (arXiv: 0807.3270v3) these IR-singularities' do not destroy the renormalization.